\documentclass[aps,prl,twocolumn,showpacs]{revtex4}
\usepackage{epsfig}
\newcommand{\be}{\begin{equation}} 
\newcommand{\ee}{\end{equation}}
\newcommand{\bea}{\begin{eqnarray}} 
\newcommand{\eea}{\end{eqnarray}}

\newcommand{\ch}{{\rm ch}} 
\newcommand{\sh}{{\rm sh}}
\renewcommand{\Re}{{\rm Re}\,\!}
\renewcommand{\Im}{{\rm Im}\,\!}
\newcommand{\C}{{ C\hspace*{-1.2ex}\rule{0.15ex}{1.5ex}\hspace*{0.9ex}}}
\newcommand{\gton}{\mathrel{\lower.9ex \hbox{$\stackrel{\displaystyle 
>}{\sim}$}}} 
\newcommand{\lton}{\mathrel{\lower.9ex \hbox{$\stackrel{\displaystyle 
<}{\sim}$}}}

\newcommand{\vp}{{\vec p}}

\newcommand{\vv}{{\vec v}}
\newcommand{\vx}{{\vec x}}

\begin{document}
\title{Elliptic flow from quark coalescence: 
mass ordering or quark number scaling?}
\author{D\'enes Moln\'ar}
\affiliation{Department of Physics, Ohio State University, 174 West 18th Ave,
Columbus, Ohio 43210, USA}

\date{\today}
\begin{abstract}
We show that either mass ordering or quark number scaling of anisotropic flow
can result from quark coalescence, depending on
the nature of phase space correlations at hadronization.
Quark number scaling  signals nonequilibrium dynamics because it
can only appear when hydrodynamic correlations break down.
However, the scaling does not hold for all nonthermal distributions,
and is compatible with covariant 
transport theory only if remarkable cancellations occur at RHIC.
\end{abstract}

\maketitle

{\em Introduction and conclusions.}
The goal of recent nuclear collision
experiments at the Relativistic Heavy Ion Collider (RHIC)
is to create extremely hot and dense nuclear matter and determine 
its properties.
An important experimental probe in this quest
is elliptic flow, $v_2 \equiv\langle \cos(2\phi)\rangle$, 
the second Fourier moment of the azimuthal momentum 
distribution~\cite{flow-review}.
Its dependence on impact parameter $b$, transverse momentum, $p_T$
and particle species provides unique constraints on the equation of 
state (EOS) and collision dynamics.

An amazing wealth of experimental $v_2(p_T)$ data is available
from $Au+Au$ at $\sqrt{s_{NN}}=130$ and $200$ GeV at RHIC 
for charged hadrons~\cite{STARv2charged,PHENIXv2charged,PHOBOSv2charged}; 
$\pi^{\pm}$, $K^{\pm}$,
and $p+\bar p$~\cite{STARv2identA,PHENIXv2identA};
$\pi^0$~\cite{PHENIXv2identB};
$K^0_S$, and $\Lambda+\bar \Lambda$~\cite{STARv2identB};
and $\Xi^-+\bar \Xi^+$ and $\Omega^-+\bar \Omega^+$~\cite{STARv2identC}.
Below $p_T \lton 1-1.5$ GeV, and for $b \lton 8$ fm,
these data can be
reproduced fairly well using ideal (Euler) 
hydrodynamics~\cite{Ollitrault,Huovinen:2001cy, Kolb:2000sd,Teaneyhydro,Hiranohydro}.
In that theory, flavor dependence enters mainly through the mass of 
the particles, and the $v_2(p_T)$ 
{\em mass ordering} pattern depends on
the EOS~\cite{Huovinen:2001cy, Kolb:2000sd,Teaneyhydro}.
Remarkably, the data indicate
a first-order 
deconfinement phase transition to the so called quark-gluon plasma (QGP) phase.

In contrast, 
a spectacular {\em quark number scaling}~\cite{Voloshincoal,coal,charmcoal} 
of elliptic flow $v_2(p_T)$ is predicted
from hadronization 
via quark coalescence~\cite{Hwacoal,Texascoal,Dukecoal,dyncoal}.
In the coalescence approach mesons(baryons) form from two(three) comoving
quarks/antiquarks,
leading to a unique meson-baryon differentiation in the flow pattern.

Though mesons tend to be lighter than baryons,
quark number scaling is in general different from mass ordering.
Comparison between precise elliptic flow $v_2(p_T)$ data for {\em heavy} mesons,
such as the $\phi(1020)$ or $K^*(892)$, and {\em light} baryons (e.g., 
protons or lambdas) could distinguish between the two behaviors.

In view of the above difference,
the realization that for thermal constituent distributions
coalescence reduces to statistical hadronization
generated significant controversy.
If, in that case, hadronization is the same both from 
coalescence and hydrodynamics, 
how could the resulting hadronic momentum anisotropies be different?

In this paper we show that there is no contradiction at all.
Either of the two scaling behaviors 
can result from quark coalescence, {\em depending on the
type of phasespace correlations present at hadronization}.
Thermal constituent distributions
imply unique coordinate-momentum correlations,
for which mass ordering follows from coalescence.
For quark number scaling to appear, departure from hydrodynamic behavior
is essential.
The growing experimental 
evidence~\cite{STARv2identB,PHENIXv2identA,STARv2identC} 
for quark number scaling
in the intermediate transverse momentum region $p_T \sim 2-5$ GeV at RHIC
is a signal of the breakdown of hydrodynamics above $p_T > 2$ GeV,
which complements and corroborates other indications,
such as the saturation of elliptic flow~\cite{STARv2charged,v2,Teaneyviscos}.

Quark number scaling emerges only under certain conditions.
Either only a subset of all possible local momentum anisotropies 
can be present, or important cancellations between contributions 
by the different anisotropies need to occur.
Only the latter possibility is compatible with covariant transport theory,
in which high-$p_T$ particles are emitted in a strongly preferred direction locally.

Here we consider momentum anisotropies right after
hadronization but before resonance decays.
Secondary production affects significantly the
final anisotropies in hydrodynamics,
but seems to influence little the quark number scaling
pattern~\cite{texasRes}.
At large $p_T \gton 5$ GeV, contributions from jet 
fragmentation (ignored here) also need to be considered.

In addition, this study considers coalescence on a 3D spacetime hypersurface.
In coalescence from diffuse 4D freezeout distributions in spacetime,
constituent spacetime and space-momentum correlations 
influence the final hadron distributions in a more complex way~\cite{dyncoal}.

{\em Anisotropic flow.}
Momentum anisotropies can be characterized via the Fourier expansion 
of the freezeout source distribution 
$S(x,\vp_T,y)\equiv dN/d^4x\, d^2p_T\, dy$
in terms of the momentum azimuthal angle $\phi$
\bea
&&\!\!\!\!\!\!\!\!\!\!S(x,\vp_T,y) 
\nonumber\\
&&\equiv 
   S_0(x,p_T,y) 
   \left[1 + 2 \sum\limits_{n=1}^\infty \Re\left(v_n(x,p_T,y)\, e^{- i n\phi}\right)\right] \ \ 
\eea
Here $\vp_T \equiv p_T (\cos \phi,\sin \phi)$,
$\phi=0$ is the impact parameter direction,
$y\equiv 0.5 \ln[(E+p_z)/(E-p_z)]$ is the rapidity,
and $v_n(x,p_T,y) \in \C$ 
is the $n$-th order {\em momentum anisotropy coefficient} for
particles emitted from an infinitesimal spacetime volume around $x$.
The real and imaginary parts of $v_n$ correspond to 
$\cos(n\phi)$ and $\sin(n\phi)$ terms.
The final observable anisotropy $\bar v_n$ is the weighted average
\be
\bar v_n(p_T,y) = \frac{\int d^4x\, S_0(x,p_T,y)\, v_n(x,p_T,y)}
             {\int d^4x\, S_0(x,p_T,y)} \ ,
\label{vneq}
\ee
which is {\em always real} due to reflection symmetry across the collision plane 
$\phi \to -\phi$ (for spherically symmetric nuclei).
In addition, for a symmetric collision system,  
all odd $\bar v_n$ vanish at midrapidity $y\approx 0$ 
due to $\phi \to (\pi-\phi)$ symmetry.
 
In the following rapidity arguments will be often suppressed for brevity.

{\em Momentum anisotropies from ideal hydrodynamics.}
Two main assumptions in approaches based on ideal hydrodynamics are
local kinetic equilibrium and 
sudden freezeout on a 3D spacetime hypersurface.
In such a case
\be
S(x,\vp) = p^\mu \sigma_\mu(x)\, 
  \delta\!\left(x^\nu \sigma_\nu(x) - x_0^\nu \sigma_\nu(x_0)\right) f(x,\vp) 
\label{HSkernel}
\ee
where $\sigma_\mu(x)$ is the normalized normal vector of the hypersurface,
while $x_0$ is an arbitrary point on the hypersurface\footnote{%
\label{HS}
Note, $\sigma\! \cdot\! x$ is a generalized time, and
$d^4 x S \equiv p\!\cdot\! d^3\sigma(x) f$.},
and 
\be
  f_{th}(x,\vp) = \frac{g}{(2\pi)^3} 
             \left\{e^{[p^\nu u_\nu(x) - \mu(x)]/T(x)}+a_s\right\}^{-1} \ ,
\label{thermalf}
\ee
where $g$ is the degeneracy factor, and $a_s=-1$, $1$, or $0$ respectively 
for bosons, fermions, or Boltzmann statistics.
$T$, $\mu$, and $u_\mu$ are the local temperature, chemical potential, and flow
velocity.

Momentum anisotropies have two sources,
the hypersurface ($p \cdot \sigma$) and the flow profile 
($p \cdot u$).
For hypersurfaces typically assumed in analytic studies,
such as constant time or constant $\tau\equiv \sqrt{t^2-z^2}$
hypersurfaces,
$p \cdot \sigma$ is independent of $\phi$.
Therefore, for simplicity we focus here on anisotropies {\em due to
the flow field only}.
Note, in general, $\phi$ dependence can appear even for azimuthally symmetric
hypersurfaces~\cite{WongHeinz}.

The anisotropy coefficients in general depend on the particle species,
mainly because of the difference in mass.
Statistics and the chemical potentials 
only play a role at very low $m_T\equiv \sqrt{m^2+p_T^2} \sim \mu,T$.
(At freezeout at RHIC, $T\sim 100-130$ MeV, while $\mu$ increases
from  $\sim 80$ MeV for pions to $\sim 350$ MeV 
for heavy particles~\cite{Hiranohydro}.)

To demonstrate the mass dependence,
consider not very low momenta, so that Boltzmann statistics is applicable.
The flow field can be characterized 
by a local transverse and longitudinal rapidity component, 
$\vv_T(x)\equiv v_T(\cos \Phi_v, \sin \Phi_v)$ and $\tilde y(x)$, 
as $u^\mu(x)\equiv(\ch \tilde y, \vv_T, \sh \tilde y) / \sqrt{1-v_T^2}$.
It is simple to show that
\be
v_n(x,p_T,y) = e^{i n\Phi_v(x)}\, 
\frac{I_n\left(\sh y_T(x)\, p_T/T(x)\right)}
     {I_0\left(\sh y_T(x)\, p_T/T(x)\right)} \ ,
\label{thermalvn}
\ee
where $\sh y_T \equiv v_T /\sqrt{1-v_T^2}$ and $\left\{I_n\right\}$ are 
the modified Bessel functions.
Remarkably, the {\em local} 
anisotropy depends only on $p_T$, and therefore is the same for all particles 
and rapidities. 
Note, at low $p_T$,
\be 
|v_n| \approx (\sh y_T\, p_T/2T)^n/n! \, \propto \, p_T^n \ ,
\label{thermalvnlowpt}
\ee
as dictated by general analyticity arguments~\cite{Danielewiczptn},
while at high $p_T$, $|v_n|\approx 1 - const \times T/ p_T \to 1$.

The coordinate-averaged flow coefficients (\ref{vneq}), 
on the other hand, in general {\em decrease in magnitude with particle mass}.
This follows from the properties of the weight $S_0$,
which via (\ref{HSkernel}) corresponds to the distribution
$$
f_0 = \exp\!\!\left[-\frac{m_T \ch(y-\tilde y(x))\ch y_T(x)}{T(x)} \right] 
      I_0\!\!\left(\frac{p_T \sh y_T(x)}{T(x)}\right)
$$
While $|v_n|$ monotonically increases with the radial flow
$y_T$, the only mass dependent part in $f_0$,
the exponential, {\em decreases} with $y_T$.
The decrease is sharper for larger mass.
Therefore for heavier particles, smaller $v_n$ values
 get preferred.
Though this argument ignores the $x$-dependence of the cosine term 
from the exponential in  (\ref{thermalvn}), 
which {\em in principle} could reverse the rise of $\Re\, v_n$ 
with $y_T$ and therefore affect the mass dependence pattern,
the general trend prevails in practice.

For more details of the derivations in the case of $\Re v_2$,
see~\cite{WongHeinz}, for example.

{\em Anisotropic flow from quark coalescence.}
In the quark coalescence approach, constituent quarks/antiquarks that are close
in phasespace can combine to form hadrons.
Assuming coalescence occurs on a 3D spacetime hypersurface,
the hadron binding energies are small,
and coalescence is a relatively rare process~\cite{coal,charmcoal},
the invariant hadron momentum distribution 
can be expressed~\cite{Oldercoal,Dukecoal,Texascoal,charmcoal}
in terms of the constituent phasespace distributions 
and the hadron Wigner functions.

Ignoring variations of phasespace distributions on length and momentum scales 
corresponding to a typical hadron ($\sim 1$ fm and $\sim 200$ MeV),
the phasespace distributions of mesons and baryons
created via $\alpha\beta \to M$ and $\alpha\beta\gamma\to B$ can be written as
\bea
 f_B(x,\vp) &=& \frac{(2\pi)^3 g_B}{g_\alpha\, g_\beta\, g_\gamma} \,
                f_\alpha(x,\vp_\alpha) \, f_\beta(x,\vp_\beta)
                f_\gamma(x,\vp_\gamma)
\nonumber \\
 f_M(x,\vp) &=& \frac{(2\pi)^6 g_M}{g_\alpha\, g_\beta} \, 
                f_\alpha(x,\vp_\alpha) \, f_\beta(x,\vp_\beta)  \ .
\label{coaleq}
\eea
Here $g$ is the degeneracy of the particle (spin and color),
while $\sum \vp_i = \vp$.
Because constituents are comoving, $\vp\, ||\, \vp_i$ and the hadron 
momentum is shared roughly in proportion to constituent mass~\cite{charmcoal}.
For hadrons composed of $u$, $d$, and $s$ quarks,
the sharing is approximately equal
because of the relatively small difference between 
$m_{u,d}\approx 0.3$ GeV and $m_{s}\approx 0.5$ GeV.
On the other hand, in hadrons that also contain 
a heavy quark, e.g., $D$ mesons or the $\Lambda_c$,
the heavy quark carries most of the momentum.

A direct implication of (\ref{coaleq}) is
\bea
v_{n,B}(x,p_T) &=&\!\! \sum\limits_{i=\alpha,\beta,\gamma} \!\!
                   v_{n,i}(x,p_{T,i}) + \Delta v_{n,B}(x, p_T)
\nonumber\\
v_{n,M}(x,p_T) &=&\!\! \sum\limits_{i=\alpha,\beta} v_{n,i}(x,p_{T,i})
                 + \Delta v_{n,M}(x,p_T)   \ ,
\label{vnscaling}
\eea
where to leading order in $v_{n,i}$, 
the corrections $\Delta v_{n,M}$ and $\Delta v_{n,B}$ are
\be
\Delta v_{n} = \!\!
  \sum\limits_{i\ne j, k=1}^\infty \!\! v_{{n+k},i} v_{k,j}^* 
  + \!\! \sum\limits_{i < j, k=1}^{n-1} \!\! v_{n-k,i} v_{k,j}
  + {\cal O}(\{v_{\ell,i}\}^3)
\label{vnscalingviol}
\ee 
with arguments identical to those in (\ref{vnscaling})
dropped for brevity. 
Thus, if the corrections are small,
$v_n$ is additive, and
in the absence of quark flavor dependence
{\em scales with quark number}~\cite{Voloshincoal,coal}.

Assuming, as for example in~\cite{coal}, 
that the spatial and momentum dependence of phasespace distributions 
{\em factorize},
the local $v_n$ coefficients are the same everywhere and therefore
the scaling carries over to the observable spacetime averages $\bar v_n$.
In this case, the local $v_n$ must of course be real,
and at RHIC the odd ones must vanish at midrapidity.
The neglect of the 
correction terms $\Delta v_2$ for elliptic flow is then justified
because the data show~\cite{STARvn} $\bar v_6 \ll \bar v_4 \ll \bar v_2 \ll 1$.

On the other hand, the scaling does not hold for arbitrary distributions.
For example, for thermal
(Boltzmann) constituent phasespace distributions,
coalescence (\ref{coaleq}) gives
thermal hadron distributions because, 
for the weak bound states assumed, $m_M \simeq m_\alpha + m_\beta$,
$m_B \simeq m_\alpha + m_\beta + m_\gamma$~\cite{Oldercoal}.
Therefore, anisotropies are the same as from hydrodynamics,
i.e., depend on the hadron mass.
This also follows from a direct calculation.
For mesons, the kernel is
$f_0 \to f_{0,\alpha} f_{0,\beta} [1+\sum\limits_{n=1}^{\infty} 
(v_{n,\alpha} v_{n,\beta}^* + c.c.)]$,
which is identical to the hydrodynamic one with $m = m_M$.
The exponential part is the same
because $p_T = p_{T,\alpha} + p_{T,\beta}$, 
momenta are shared such that $m_T \simeq m_{T,\alpha} + m_{T,\beta}$,
and therefore also $y\simeq y_\alpha \simeq y_\beta$.
The rest of the kernel and all meson flow anisotropies also agree,
as can be shown using the {\em exact} flow addition formula
\bea
&&\!\!\!\!\!\!\!\!\!\!v_{n,M} = 
\nonumber\\
&=&\frac{\sum\limits_{i=\alpha,\beta}\!\!v_{n,i}
      + \sum\limits_{k=1}^\infty \! (v_{{n+k},\alpha} v_{k,\beta}^* 
                                     + \alpha \leftrightarrow \beta) 
      + \!\! \sum\limits_{k=1}^{n-1} \!\! v_{n-k,\alpha} v_{k,\beta}}
     {1 + \sum\limits_{k=1}^\infty (v_{k,\alpha} v_{k,\beta}^* + c.c.) }
\nonumber\\
\label{exactMvn}
\eea
and the Bessel function addition theorem. The direct proof
for baryons is analogous but more involved.

Thus, for thermal constituent distributions, momentum anisotropies from 
quark coalescence {\em depend on mass, and {\em not} quark number}.
There is an intuitive way to see why this must be so. 
In the frame where the flow velocity is zero,
the momentum distribution is isotropic for {\em all} constituents.
Therefore, in that frame all $v_n$ vanish, and hence (\ref{vnscaling})
gives no hadron anisotropies.
The only source of anisotropy then is the 
Lorentz boost back to the laboratory frame,
which depends only on the particle mass, momentum, and the boost velocity.
{\em This is also true even if hydrodynamics breaks down,
as long as distributions are of the form $f(x,\vp)=g(p\cdot u(x),x)$.}
Quark number scaling can emerge only if,
in any frame,
at least one of the constituent distributions is anisotropic.

In the thermal case, quark number scaling is
violated because the nonlinear terms (\ref{vnscalingviol}) are important.
Consider for example $v_2$, at 
low $p_T$ so that (\ref{thermalvnlowpt}) is justified.
To leading ${\cal O}(p_T^2)$ order, both the linear and 
the $v_1 v_1$ terms contribute. In fact the latter give
as much as half the meson $v_2$, and two-thirds for baryons.
For higher-order flow coefficients, all $v_{n-k} v_k$ terms contribute at
leading order in $p_T$.

Therefore, one class of freezeout distributions 
(besides uniform local anisotropy $v_n$) for which the observable
$\bar v_n$ scales with quark number is
when \be
|v_n| \gg |v_{n-k}| |v_k|
\label{linscaling}
\ee  
in the spacetime region where {\em most} particles are emitted from.
In this case, there is (approximate) scaling locally and, 
because the formula is linear, 
scaling is preserved upon averaging.
Note, a {\em small} amount of other anisotropies, for example 
$v_1 \sim v_2^{3/2}$ besides a pure $v_2$,
can even help compensate
the denominator in (\ref{exactMvn}), which otherwise tends to reduce
the flow relative to the scaling expectation.

\begin{figure}[htpb]
\centerline{\epsfig{file=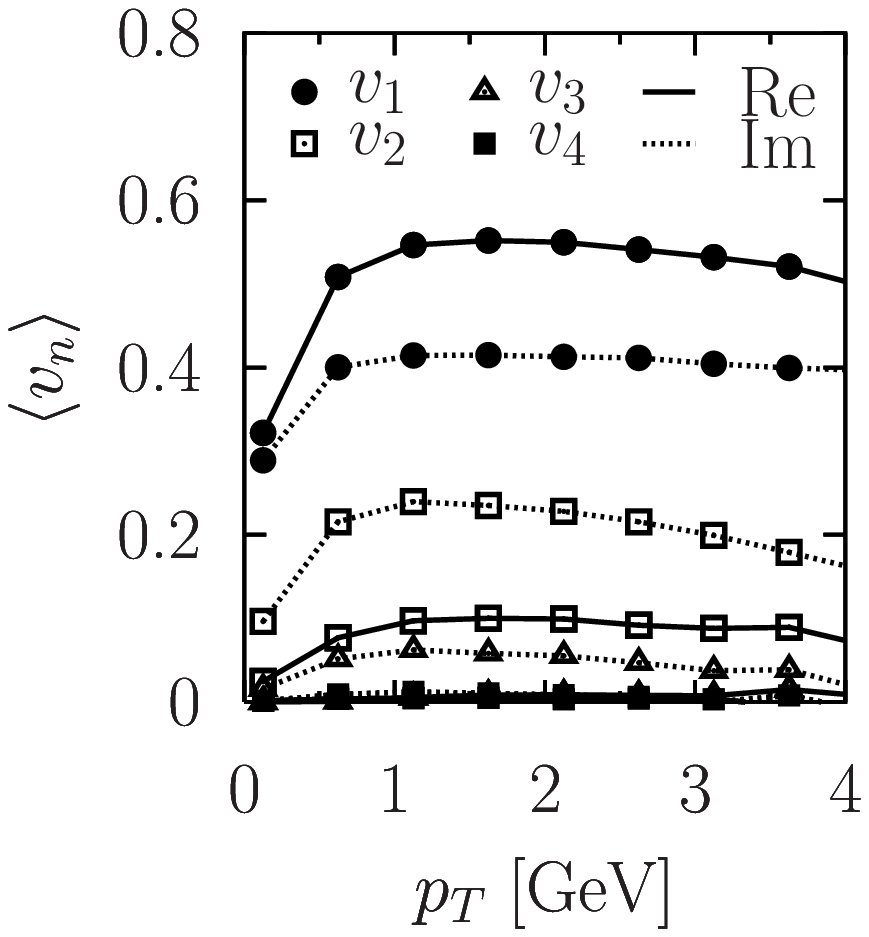,width=1.65in,height=1.8in,angle=0}%
\hskip 0.2cm
\epsfig{file=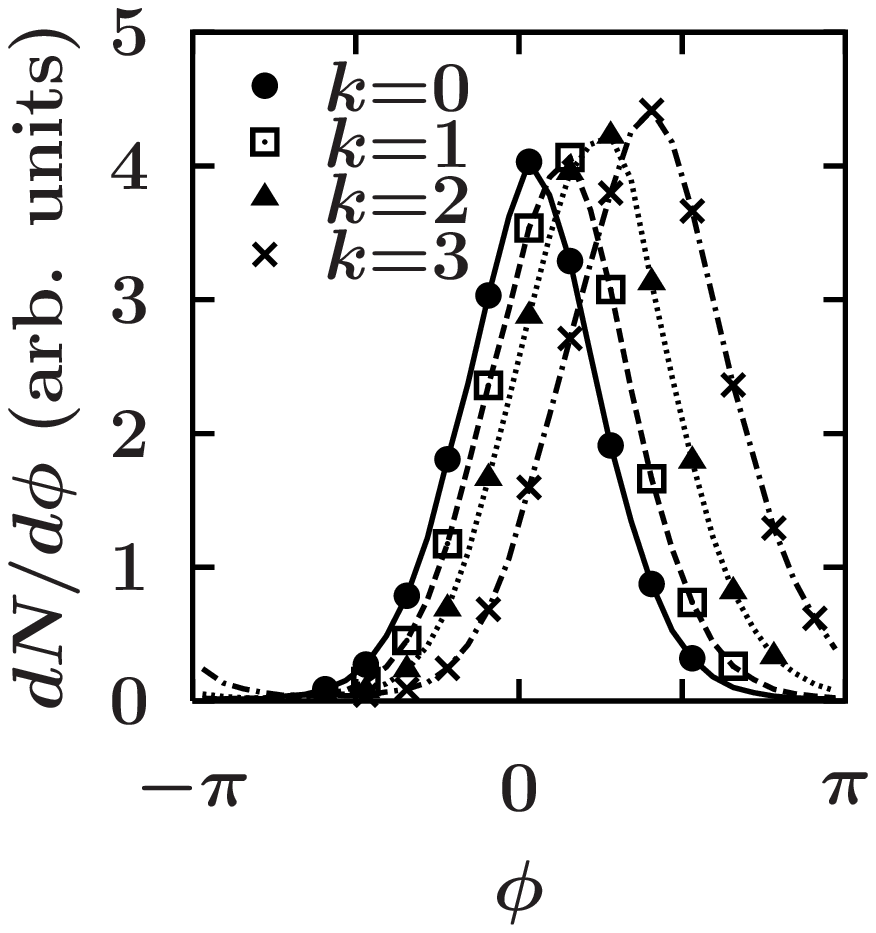,width=1.7in,height=1.8in,angle=0}}
\caption{Results from 
MPC~\cite{MPC,v2} for $Au+Au$ at $\sqrt{s_{NN}}=200$ GeV with $b=8$ fm. 
Left: 
local constituent quark anisotropies at midrapidity as a function of $p_T$,
averaged over the first quadrant of the transverse plane. 
Right:
$\phi$ distributions at midrapidity for $2<p_T<3$ GeV,
averaged over four wedges in the transverse plane 
$\varphi_x \in [k\pi/8, (k+1)\pi/8]$, with $k = 0...3$.}
\label{fig:1}
\end{figure}

The scaling may also hold even
if the nonlinear terms are important. 
However, that requires a high degree of fortuitous cancellations 
at RHIC,
as illustrated in Fig.~\ref{fig:1}.
Results from the covariant parton transport model 
MPC~\cite{MPC,v2,dyncoal} are shown for $Au+Au$ at
$\sqrt{s_{NN}}=200$ GeV with $b=8$ fm,
from a calculation identical to the one in~\cite{dyncoal} 
with a gluon-gluon cross section $\sigma_{gg}=10$~mb.
In the left panel local constituent quark anisotropies up to fourth order
are plotted as a function of transverse momentum, 
averaged over the first quadrant of the transverse plane 
($0 < \varphi_x < \pi/2$, where 
$\vx_T \equiv x_T (\cos \varphi_x, \sin \varphi_x)$).
Coefficients that vanish upon a {\em full} spacetime average due to symmetry, 
in particular $v_1$ and $\Im v_2$, are surprisingly large.
Therefore, (\ref{linscaling}) does not hold and, for example,
the denominator in (\ref{exactMvn}) exceeds 2 above $p_{T,i} \gton 1$ GeV.

The origin of these large anisotropies
is shown in the right panel of Fig.~\ref{fig:1},
where we plot the $\phi$ distributions at midrapidity for $2<p_T<3$ GeV,
averaged over four wedges in the transverse plane
$\varphi_x \in [k\pi/8, (k+1)\pi/8]$, with $k = 0,1,2,3$.
Instead of small harmonic modulations over a uniform background,
the distributions are strongly peaked
because high-$p_T$ particles can only escape from a surface layer of the 
reaction region. 
In this case, anisotropies from coalescence change their character
completely.
Consider, for simplicity, Gaussian constituent distributions
around an azimuthal direction $\phi_0$,
$S_i(x,\vp_T,y)=C_i(x,p_T,y) \exp[-(\phi-\phi_0)^2/(2\sigma_i^2)]$.
For not too wide $\sigma\lton 2/\sqrt{n+1}$, 
the constituent and hadron anisotropies
\be
v_{n,i} \simeq e^{in\phi_0} e^{-n^2\sigma_i^2/2}, \ \ 
v_{n,h} \simeq e^{in\phi_0} e^{-n^2/(2 \sum_i \sigma_i^{-2})},
\ee
scale as $(const)^{-n^2}$. This is very different from both (\ref{thermalvn})
and  (\ref{linscaling}).
For example, if all constituents have the same width $\sigma$, we have
$$
|v_{n,B}(3 p_T)| \simeq |v_{n,q}(p_T)|^{1/3}, \ \ 
|v_{n,M}(2 p_T)| \simeq |v_{n,q}(p_T)|^{1/2},
$$
which gives for $|v_{n,q}| \gton 0.1$ smaller baryon/meson flow ratios,
and for $|v_{n,q}| > 0.25$ also smaller hadron flows,
than $v_{n,B}(3 p_T) = 3v_{n,q}(p_T)$, $v_{n,M}(2p_T) = 2 v_{n,q}(p_T)$
from linear scaling (\ref{vnscaling}).

The observable averages $\{\bar v_n\}$ 
are, of course, determined by the interplay between variations in 
the local emission angle, the width, and constituent density.
It is quite remarkable that the end result 
in a dynamical coalescence
approach~\cite{dyncoal} is only 
a modest $\sim 20$\% and $\sim 30$\% reduction of
pion and proton elliptic flow at RHIC relative to quark number scaling.

The above results underscore the importance of {\em local} momentum 
anisotropies in hadronization via quark coalescence.
Flow addition formulas apply locally, even though only 
spacetime averages of the final anisotropies can be observed.
Odd-order $\{v_n\}$ and the sine terms play a major role, even at midrapidity.
These terms have been ignored in analytic studies so far~\cite{coal,Kolbcoal}
but are naturally incorporated 
in a dynamical approach~\cite{dyncoal}.

{\em Acknowledgments.}
Helpful discussions with E.~Shuryak and U.~Heinz are gratefully acknowledged. 
This work was supported by DOE grant DE-FG02-01ER41190.


\begin{thebibliography}{99}

\bibitem{flow-review}
For reviews see, e.g., 
J.~Ollitrault, Nucl. Phys. A {\bf 638}, 195 (1998);
%\cite{Poskanzer:2001cx}
%\bibitem{Poskanzer:2001cx}
A.~M.~Poskanzer,
%``Anisotropic flow at the SPS and RHIC,''
nucl-ex/0110013;
or 
%\bibitem{Voloshin:2002wa}
S.~A.~Voloshin,
%``Anisotropic flow,''
Nucl.\ Phys.\ A {\bf 715}, 379 (2003).
%[arXiv:nucl-ex/0210014].
%%CITATION = NUCL-EX 0210014;%%

\bibitem{STARv2charged}
C.~Adler {\it et al.}  [STAR Collaboration],
%``Azimuthal anisotropy and correlations in the hard scattering regime at  RHIC''
Phys.\ Rev.\ Lett.\  {\bf 90}, 032301 (2003);
%%CITATION = NUCL-EX 0206006;%%
% --------- pi, K, K0, p, lambda 
K.~Filimonov  [STAR Collaboration],
%``Azimuthal Anisotropy of Charged and Identified High $p_T$ Hadrons in Au+Au Collisions at RHIC,''
Nucl.\ Phys.\ A{\bf 715}, 737 (2003).

\bibitem{PHENIXv2charged}
%\cite{Adcox:2002ms}
%\bibitem{Adcox:2002ms}
K.~Adcox {\it et al.}  [PHENIX Collaboration],
%``Flow measurements via two-particle azimuthal correlations in Au + Au
%collisions at s(NN)**(1/2) = 130-GeV,''
Phys.\ Rev.\ Lett.\  {\bf 89}, 212301 (2002);
%[arXiv:nucl-ex/0204005].
%%CITATION = NUCL-EX 0204005;%%
%\cite{Ajitanand:2002qd}
%\bibitem{Ajitanand:2002qd}
N.~N.~Ajitanand  [PHENIX Collaboration],
%``Two particle azimuthal correlation measurements in PHENIX,''
Nucl.\ Phys.\ A {\bf 715}, 765 (2003).
%[arXiv:nucl-ex/0210007].
%%CITATION = NUCL-EX 0210007;%%

\bibitem{PHOBOSv2charged}
%\cite{unknown:2004mh}
%\bibitem{unknown:2004mh}
B.~B.~Back {\it et al.}
[Phobos Collaboration],
%``Centrality and pseudorapidity dependence of elliptic flow for charged hadrons
%in Au + Au collisions at s(NN)**(1/2) = 200-GeV,''
nucl-ex/0407012.
%%CITATION = NUCL-EX 0407012;%%

\bibitem{STARv2identA} %pi, K, p
%\cite{Adler:2001nb}
%\bibitem{Adler:2001nb}
C.~Adler {\it et al.}  [STAR Collaboration],
%``Identified particle elliptic flow in Au + Au collisions at  s(NN)**(1/2) =
%130-GeV,''
Phys.\ Rev.\ Lett.\  {\bf 87}, 182301 (2001).
%[arXiv:nucl-ex/0107003].
%%CITATION = NUCL-EX 0107003;%%

\bibitem{PHENIXv2identA} %pi,K,p
%\bibitem{Adler:2003kt}
S.~S.~Adler {\it et al.}  [PHENIX Collaboration],
%``Elliptic flow of identified hadrons in Au + Au collisions at  s(NN)**(1/2) =
%200-GeV,''
Phys.\ Rev.\ Lett.\  {\bf 91}, 182301 (2003).
%[arXiv:nucl-ex/0305013].
%%CITATION = NUCL-EX 0305013;%%


\bibitem{PHENIXv2identB} %pi0, gamma
%\cite{Kaneta:2004sb}
%\bibitem{Kaneta:2004sb}
M.~Kaneta  [PHENIX Collaboration],
%``Event anisotropy of identified pi0, photon and electron compared to charged
%pi, K, p and deuteron in s(NN)**(1/2) = 200-GeV Au + Au at PHENIX,''
nucl-ex/0404014.
%%CITATION = NUCL-EX 0404014;%%

\bibitem{STARv2identB} %K0, lambda
%\cite{Adams:2003am}
%\bibitem{Adams:2003am}
J.~Adams {\it et al.}  [STAR Collaboration],
%``Particle dependence of azimuthal anisotropy and nuclear modification of
%particle production at moderate p(T) in Au + Au collisions at  s(NN)**(1/2) =
%200-GeV,''
Phys.\ Rev.\ Lett.\  {\bf 92}, 052302 (2004);
%[arXiv:nucl-ex/0306007].
%%CITATION = NUCL-EX 0306007;%%
%\cite{Adler:2002pb}
%\bibitem{Adler:2002pb}
C.~Adler {\it et al.}  [STAR Collaboration],
%``Azimuthal anisotropy of K0(S) and Lambda + anti-Lambda production at
%mid-rapidity from Au + Au collisions at s(NN)**(1/2) = 130-GeV,''
Phys.\ Rev.\ Lett.\  {\bf 89}, 132301 (2002).
%[arXiv:hep-ex/0205072].
%%CITATION = HEP-EX 0205072;%%

\bibitem{STARv2identC} %xi, omega
%\cite{Castillo:2004jy}
%\bibitem{Castillo:2004jy}
J.~Castillo  [STAR Collaboration],
%``Elliptic flow of multi-strange baryons Xi and Omega in Au + Au collisions at
%s(NN)**(1/2) = 200-GeV,''
nucl-ex/0403027.

\bibitem{Ollitrault}
J.~Ollitrault,
%``Anisotropy as a signature of transverse collective flow,''
Phys.\ Rev.\ {\bf D 46}, 229 (1992).

\bibitem{Kolb:2000sd}
P.~F.~Kolb, J.~Sollfrank and U.~Heinz,
%``Anisotropic transverse flow and the quark-hadron phase transition,''
Phys.\ Rev.\ C {\bf 62}, 054909 (2000).
%[arXiv:hep-ph/0006129].
%%CITATION = HEP-PH 0006129;%%

\bibitem{Huovinen:2001cy}
P.~Huovinen {\it et al.},
%, P.~F.~Kolb, U.~Heinz, P.~V.~Ruuskanen and S.~A.~Voloshin,
%``Radial and elliptic flow at RHIC: Further predictions,''
Phys.\ Lett.\ B {\bf 503}, 58 (2001).
%[hep-ph/0101136].
%%CITATION = HEP-PH 0101136;%%

\bibitem{Teaneyhydro}
D.~Teaney, J.~Lauret and E.~V.~Shuryak,
%``A hydrodynamic description of heavy ion collisions at the SPS and RHIC,''
nucl-th/0110037.
%%CITATION = NUCL-TH 0110037;%%

\bibitem{Hiranohydro}
%\cite{Hirano:2002ds}
%\bibitem{Hirano:2002ds}
T.~Hirano and K.~Tsuda,
%``Collective flow and two pion correlations from a relativistic  hydrodynamic
%model with early chemical freeze out,''
Phys.\ Rev.\ C {\bf 66}, 054905 (2002);
%[arXiv:nucl-th/0205043].
%%CITATION = NUCL-TH 0205043;%%
T.~Hirano, private communication.

\bibitem{Voloshincoal}
S.~A.~Voloshin,
%``Anisotropic flow,''
{\it Nucl.\ Phys.\ }A {\bf 715} (2003) 379.
%[arXiv:nucl-ex/0210014].
%%CITATION = NUCL-EX 0210014;%%

%\cite{Molnar:2003ff}
\bibitem{coal}
%\bibitem{Molnar:2003ff}
D.~Molnar and S.~A.~Voloshin,
%``Elliptic flow at large transverse momenta from quark coalescence,''
Phys.\ Rev.\ Lett.\  {\bf 91}, 092301 (2003).
%[arXiv:nucl-th/0302014].
%%CITATION = NUCL-TH 0302014;%%

\bibitem{charmcoal}
%\cite{Lin:2003jy}
%\bibitem{Lin:2003jy}
Z.~w.~Lin and D.~Molnar,
%``Quark coalescence and elliptic flow of charm hadrons,''
{\it Phys.\ Rev.\ }C {\bf 68} (2003) 044901.
%%CITATION = NUCL-TH 0304045;%%

\bibitem{Dukecoal}
%\cite{Fries:2003vb}
%R.~J.~Fries, S.~A.~Bass, B.~Muller and C.~Nonaka,
R.~J.~Fries {\it et al.},
%``Hadronization in heavy ion collisions: Recombination and fragmentation  of partons,''
Phys.\ Rev.\ Lett.\  {\bf 90}, 202303 (2003);
%nucl-th/0301087.
%%CITATION = NUCL-TH 0301087;%%
%\cite{Fries:2003kq}
%\bibitem{Fries:2003kq}
%R.~J.~Fries, B.~Muller, C.~Nonaka and S.~A.~Bass,
R.~J.~Fries {\it et al.},
%``Hadron production in heavy ion collisions: Fragmentation and  recombination
%from a dense parton phase,''
Phys.\ Rev.\ C {\bf 68}, 044902 (2003).
%[arXiv:nucl-th/0306027].
%%CITATION = NUCL-TH 0306027;%%

\bibitem{Texascoal}
%\cite{Greco:2003xt}
V.~Greco, C.~M.~Ko and P.~Levai,
%``Parton coalescence and antiproton/pion anomaly at RHIC,''
%nucl-th/0301093.
Phys.\ Rev.\ Lett.\  {\bf 90}, 202302 (2003);
%%CITATION = NUCL-TH 0301093;%%
%\cite{Greco:2003mm}
%\bibitem{Greco:2003mm}
%V.~Greco, C.~M.~Ko and P.~Levai,
%``Parton coalescence at RHIC,''
Phys.\ Rev.\ C {\bf 68}, 034904 (2003).
%[arXiv:nucl-th/0305024].
%%CITATION = NUCL-TH 0305024;%%

\bibitem{Hwacoal}
%\bibitem{Hwa:2002zu}
R.~C.~Hwa and C.~B.~Yang,
%``Inclusive distributions for hadronic collisions in the valon  recombination model,''
Phys.\ Rev.\ C {\bf 66}, 025205 (2002); 
%[arXiv:hep-ph/0204289].
%%CITATION = HEP-PH 0204289;%%
%\bibitem{Hwa:2002tu}
%``Scaling behavior at high p(T) and the p/pi ratio,''
{\it ibid.} {\bf 67}, 034902 (2003).
%[arXiv:nucl-th/0211010].
%%CITATION = NUCL-TH 0211010;%%

\bibitem{dyncoal}
D.~Molnar, nucl-th/0406066;
%\cite{Molnar:2004ei}
%\bibitem{Molnar:2004ei}
%D.~Molnar,
%``Particle correlations at RHIC from parton coalescence dynamics: First
%results,''
J.\ Phys.\ G {\bf 30}, S1239 (2004).
%[arXiv:nucl-th/0403035].
%%CITATION = NUCL-TH 0403035;%%

\bibitem{v2}
D.~Molnar and M.~Gyulassy,
%``Saturation of elliptic flow at RHIC: Results from the covariant elastic  parton cascade model MPC,''
Nucl.\ Phys. {\bf A697}, 495 (2002), {\bf A703}, 893(E) (2002);
%[arXiv:nucl-th/0104073].
%%CITATION = NUCL-TH 0104073;%%
%``Elliptic flow and freeze-out from the parton cascade MPC,''
{\it ibid.} {\bf A698}, 379 (2002).
%[arXiv:nucl-th/0104018]; 
%%CITATION = NUCL-TH 0104018;%%

\bibitem{Teaneyviscos} D.~Teaney,
%``Viscous Corrections to Spectra, Elliptic Flow, and HBT Radii,''
Nucl.\ Phys. {\bf A715}, 817 (2003);
%``Effect of shear viscosity on spectra, elliptic flow, and Hanbury Brown-Twiss
% radii''
Phys.\ Rev. {\bf C68}, 034913 (2003).

\bibitem{texasRes}
%\cite{Greco:2004ex}
%\bibitem{Greco:2004ex}
V.~Greco and C.~M.~Ko,
%``Effect of resonance decays on hadron elliptic flows,''
nucl-th/0402020.
%%CITATION = NUCL-TH 0402020;%%

\bibitem{WongHeinz}
%\cite{Heinz:2002rs}
%\bibitem{Heinz:2002rs}
U.~Heinz and S.~M.~H.~Wong,
%``Elliptic flow from a transversally thermalized fireball,''
Phys.\ Rev.\ C {\bf 66}, 014907 (2002).
%[arXiv:hep-ph/0205058].
%%CITATION = HEP-PH 0205058;%%

\bibitem{Danielewiczptn} P.~Danielewicz,
%``''Effects of compression and collective expansion on particle emission
%from central heavy-ion reactions''
Phys.\ Rev.\ C {\bf 51}, 716 (1995).

\bibitem{Oldercoal}
%\bibitem{Dover:1991zn}
C.~B.~Dover {\it et al.},
%C.~B.~Dover, U.~W.~Heinz, E.~Schnedermann and J.~Zimanyi,
%``Relativistic coalescence model for high-energy nuclear collisions,''
Phys.\ Rev.\ C {\bf 44}, 1636 (1991);
%%CITATION = PHRVA,C44,1636;%%
%\cite{Scheibl:1998tk}
%\bibitem{Scheibl:1998tk}
R.~Scheibl and U.~Heinz,
%``Coalescence and flow in ultra-relativistic heavy ion collisions,''
Phys.\ Rev.\ C {\bf 59}, 1585 (1999).
%[arXiv:nucl-th/9809092].
%%CITATION = NUCL-TH 9809092;%%

\bibitem{STARvn} % v1, v2, v4, v6
%\cite{Adams:2003zg}
%\bibitem{Adams:2003zg}
J.~Adams {\it et al.}  [STAR Collaboration],
%``Azimuthal anisotropy at RHIC: The first and fourth harmonics,''
Phys.\ Rev.\ Lett.\  {\bf 92}, 062301 (2004).
%[arXiv:nucl-ex/0310029].
%%CITATION = NUCL-EX 0310029;%%

\bibitem{MPC} D.~Moln\'ar, MPC~1.6.7.
This parton cascade code is available at
http://nt3.phys.columbia.edu/people/molnard.

\bibitem{Kolbcoal}
%\cite{Kolb:2004gi}
%\bibitem{Kolb:2004gi}
%P.~F.~Kolb, L.~W.~Chen, V.~Greco and C.~M.~Ko,
P.~F.~Kolb {\it et al.},
%``Momentum anisotropies in the quark coalescence model,''
Phys.\ Rev.\ C {\bf 69}, 051901 (2004).
%[arXiv:nucl-th/0402049].
%%CITATION = NUCL-TH 0402049;%%

\end{thebibliography}
\end{document}